\documentclass[aps,11pt,onecolumn,preprintnumbers,amsmath,amssymb]{revtex4}
\usepackage{amssymb}
\usepackage{amsmath}
\usepackage{graphicx}
\usepackage{bm}
\usepackage{color}

\makeatletter
\@ifundefined{textcolor}{}
{%
 \definecolor{BLACK}{gray}{0}
 \definecolor{WHITE}{gray}{1}
 \definecolor{RED}{rgb}{1,0,0}
 \definecolor{GREEN}{rgb}{0,1,0}
 \definecolor{BLUE}{rgb}{0,0,1}
 \definecolor{CYAN}{cmyk}{1,0,0,0}
 \definecolor{MAGENTA}{cmyk}{0,1,0,0}
 \definecolor{YELLOW}{cmyk}{0,0,1,0}
 }

\begin{document}
\title{Giant paramagnetism induced valley polarization of electrons in charge-tunable monolayer MoSe$_2$}
\author{Patrick Back}
\author{Meinrad Sidler}
\author{Ovidiu Cotlet}
\author{Ajit Srivastava}
\author{Naotomo Takemura}
\author{Martin Kroner}
\author{Atac Imamo\u{g}lu}

\affiliation{Institute of Quantum Electronics, ETH Z\"{u}rich,
CH-8093 Z\"{u}rich, Switzerland}

\date{\today }

\maketitle

\paragraph*{} 
\textbf{For applications exploiting the valley pseudospin degree of
freedom in transition metal dichalcogenide monolayers, efficient
preparation of electrons or holes in a single valley is essential.
Here, we show that a magnetic field of $7$ Tesla leads to a
near-complete valley polarization of electrons in MoSe$_2$ monolayer
with a density $\bm{1.6 \times 10^{12}}$ cm$\bm{^{-2}}$; in the
absence of exchange interactions favoring single-valley occupancy, a
similar degree of valley polarization would have required a
pseudospin g-factor exceeding 40. To investigate the magnetic
response, we use polarization resolved photoluminescence as well as
resonant reflection measurements. In the latter, we observe gate
voltage dependent transfer of oscillator strength from the exciton
to the attractive-Fermi-polaron: stark differences in the spectrum
of the two light helicities provide a confirmation of valley
polarization. Our findings suggest an interaction induced giant
paramagnetic response of MoSe$_2$, which paves the way for
valleytronics applications.}

\paragraph*{} %

\vspace{5 mm}


Transition metal dichalcogenide (TMD) monolayers such as MoSe$_2$
represent a new class of two dimensional (2D) direct band-gap
semiconductors~\cite{RadisavljevicNNano2011,SplendianiNanoLett2010,BaugherNNano2014,NovoselovScience2013}
exhibiting an ultra-large exciton binding energy $E_{exc}$ in the
order of
$0.5$~eV~\cite{ChernikovPRL2014,YeNature2014,HePRL2014,QiuPRL2013,WangPRL2014}
and finite Berry curvature that leads to valley Hall
effect~\cite{MakScience2014,LeeNNano2016} as well as a modification
of the exciton spectrum~\cite{SrivastavaPRL2015,ZhouPRL2015}.
Investigation of one of the most interesting features of this
material system, namely the valley pseudospin degree of
freedom~\cite{YeNPhys2016,YangNatPhys2015}, has been hampered by the
difficulty in obtaining a high-degree of valley polarization of free
electrons or holes~\cite{SanchezArxiv2016}. While circularly
polarized excitation ensures that the excitons are generated in a
single valley~\cite{CaoNComm2012,ZengNNano2012,MakNNano2012},
significant transfer of valley polarization from excitons to
itinerant electrons or holes has not been observed.

Here, we report a strong paramagnetic response of a two dimensional
electron system (2DES) in a charge-tunable monolayer MoSe$_2$
sandwiched between two hexagonal boron-nitride (hBN) layers
(Fig.~1A). Figure~1B shows the corresponding single-particle
energy-band diagram when an external magnetic field $B_z$ is applied
along the direction perpendicular to the plane of the monolayer,
lifting the degeneracy of the electronic states in the $\pm K$
valleys. Remarkably, our experiments demonstrate that while the
model depicted in Fig.~1B is qualitatively correct, the modest
electron and exciton valley Zeeman splitting predicted by
calculations neglecting interaction effects~\cite{Durnev-arxiv2016}
fails dramatically to explain the high-degree of valley polarization
we observe for a 2DES with an electron density $n_e = 1.6 \times
10^{12}$ cm$^{-2}$ at $|B| = 7$ T. Concurrently, we find that the
Zeeman splitting of elementary optical excitations out of a 2DES can
be strongly modified by interaction and phase-space filling effects,
yielding effective exciton-polaron g-factors as high as $18$.

\vspace{5 mm}


To characterize the $n_e$ dependence of the optical response at $B_z
= 7$~Tesla, we first carried out polarization resolved
photoluminescence (PL) experiments. Figure~2A and 2B show the PL
spectrum of the MoSe$_2$ monolayer as a function of the gate voltage
$V_g$ for $\sigma^+$ and $\sigma^-$ polarized emission upon
excitation with a linearly polarized excitation laser at wavelength
$\lambda_L = 719$~nm. For $V_g \ge V_{on,-K} = 100$~V  the monolayer
is devoid of free electrons ($n_e \simeq 0$). In this regime PL,
which we attribute to radiative recombination of excitons bound to
localized electrons (i.e. localized trions), exhibits a sizeable
degree of circular polarization where the ratio of the maximum peak
intensities of $\sigma^+$ and $\sigma^-$ polarized emission is
$R_{PL} \simeq 11$ (Fig.~2C). We attribute the observed PL
polarization, which vanishes completely at $B_z=0$, to fast
relaxation into the lowest energy optically excited states. As free
electrons are injected into the sample ($V_g < V_{on,-K}$), we
observe a dramatic increase (decrease) in $\sigma^+$ ($\sigma^-$)
polarized PL (Fig.~2D). The maximum value of $R_{PL}$ in this regime
exceeds $700$; the suppression of trion PL at $\lambda = 760$~nm is
so strong that in this $V_g$ range the exciton emission at $\lambda
= 748$~nm is the dominant source of $\sigma^-$ polarized photons.
Further increase of $n_e$ ($V_g < V_{on,K} = 70$~V) results in a
strong red-shift of emission as well as a recovery of $R_{PL}$ to
the value observed in the absence of free electrons (Fig~2E).

The highly nontrivial $V_g$ dependence of polarization resolved
trion PL can be understood by recalling that in MoSe$_2$ monolayers
at low $n_e$, $\sigma^+$ ($\sigma^-$) trions are formed only when a
$-K$ ($K$) valley electron binds to a $K$ ($-K$) valley exciton. A
strong increase (decrease) in the observed $\sigma^+$ ($\sigma^-$)
trion emission therefore indicates that the electrons in the range
$V_{on,K} < V_g < V_{on,-K}$, corresponding to free electron density
range $0 < n_e < 1.6 \times 10^{12}$~$cm{^{-2}}$, exhibit a high
degree of valley polarization. The fact that $\sigma^-$ exciton PL
is not reduced as $V_g$ is tuned below $V_{on,-K}$ corroborates this
conclusion.

To verify the conclusions we draw from PL experiments, we measure
white light reflection as a function of $V_g$. Figure~3A~(3B) shows
the normalized reflection spectrum of $\sigma^+$ ($\sigma^-$)
polarized white light at $B_z = 7$~Tesla: in agreement with recent
experiments~\cite{SidlerArxiv2016}, the spectrum exhibits a dominant
exciton line at vanishing mobile electron densities $n_e \sim 0$. As
free electrons are introduced into the monolayer, a red-shifted
attractive polaron resonance becomes prominent. Increasing the $n_e$
further results in a sharp blue-shift and broadening of the exciton
line, which has been identified as the repulsive
polaron~\cite{SidlerArxiv2016,Macdonald2016}. The interaction
between excitons and electrons that lead to polaron formation can be
described as being associated with a trion
channel~\cite{SidlerArxiv2016}. As a consequence, $\sigma^+$
($\sigma^-$) attractive polaron resonance is observed if and only if
the $-K$ ($K$) valley has a finite electron density, which in turn,
as stated earlier, ensures that $\sigma^+$ ($\sigma^-$) trions could
be formed.

The $\sigma^+$ and $\sigma^-$ polarized reflection spectra exhibit
two striking differences in the  range  $V_{on,K} < V_g <
V_{on,-K}$. First, the reflection spectrum of $\sigma^+$ shows an
attractive polaron resonance whereas that of $\sigma^-$ does not --
indicating that electrons predominantly occupy $-K$ valley states.
Second, while the $\sigma^+$ attractive polaron exhibits a red-shift
with increasing $n_e$, $\sigma^-$ exciton resonance shows a
blue-shift even though a corresponding $\sigma^-$ attractive polaron
resonance is absent. If there is strong electron pseudospin
polarization in -K valley, we would expect the -K valley
($\sigma^-$) exciton resonance to exhibit a blue-shift due to
phase-space filling stemming from the Pauli blocking of the
electronic states that would otherwise contribute to exciton
formation. The red-shift of the attractive polaron is in turn fully
consistent with the absence of a 2DES and the associated phase-space
filling in the $K$ valley. We therefore conclude that the reflection
spectra in this $V_g$ range is fully consistent with near-complete
valley polarization of electrons.

We model our structure as a parallel plate capacitor to determine
the change in electron density as $V_g$ is decreased from
$V_{on,-K}$ to $V_{on,K}$:
\[ \Delta n_{e} = (V_{on,-K} - V_{on,K})
\varepsilon_0 \epsilon /(e L) = 1.6 \times 10^{12} cm^{-2},\] where
$\varepsilon_0$, $e$ and $L/\epsilon = 101$~nm denote the vacuum
permittivity, the electron charge and the effective combined
thickness of the insulating SiO$_2$ and hBN layers separating the
MoSe$_2$ flake from the back gate.
Since for $V_{on,-K} > V_g > V_{on,K}$, electrons only occupy states
in $-K$ valley, $\Delta n_{e}$ gives the maximum fully valley
polarized electron density. We estimate an uncertainty of $0.5
\times 10^{12} cm^{-2}$ for the absolute value of $\Delta n_{e}$;
this uncertainty stems from the accuracy of our measurement of the
thickness of the hBN layer, as well as our inability to determine
$V_{on,\pm K}$ precisely.

We remark that reported theoretical predictions of single-particle
electron valley g-factors vary from $-0.86$~\cite{Kormanyos2016} to
$5.12$~\cite{Durnev-arxiv2016}; if we were to take the latter value,
we would obtain $\Delta n_{e} = 0.2 \times 10^{12}$~cm$^{-2}$ at
$B_z = 7$~T. The corresponding value that our experiments yield is a
factor of 8 larger, demonstrating that a simple paramagnetic
response based on single-particle parameters cannot describe the
inferred degree of valley polarization. We also note that the
$B_z$-dependence of valley polarized $n_e$ shows saturation for
$|B_z| \ge 5$~T, indicating a
deviation from a purely paramagnetic response that is consistent
with super-paramagnetism~\cite{Fonseca2002}. We speculate that
reduced screening and relatively heavy electron mass may ensure that
exchange and correlation energies in monolayer MoSe$_2$ exceed
kinetic energy even for densities of order $1.6 \times
10^{12}$~$cm^{-2}$, resulting in the observed  giant paramagnetic
response at $T= 4$~K. Investigation of higher quality samples at
lower temperatures could be used to investigate if an interaction
induced phase transition to a ferromagnetic state is
possible~\cite{ShamPRL1979}.

\vspace{5 mm}


The reflection spectra depicted in Fig.~3A,B reveal that the valley
Zeeman splitting of the excitonic transitions can be drastically
modified when $n_e > 0$. Figure~3C shows overlayed line-cuts through
the normalized reflection spectra for $V_g = 127\ V > V_{on,-K}$
($n_e \simeq 0$). The blue-shift of the $\sigma^-$ exciton line with
respect to the $\sigma^+$ exciton stems from the valley Zeeman
effect and the extracted exciton g-factor $g_{exc} = 4.4$ is in
excellent agreement with previous
reports~\cite{MacNeillPRL2015,SrivastavaNPhys2015,AivazianNPhys2015,LiPRL2014}.

Figure~3D shows the reflection spectra for $V_g = 69$~V ($n_e = 1.7
\times 10^{12}$ cm$^{-2}$). The $\sigma^+$ reflection spectrum for
this $V_g$ is dominated by the attractive polaron with a smaller
weight on the repulsive polaron/exciton branch, whereas the opposite
is true for $\sigma^-$ spectrum. An estimation of the
peak-splittings for the attractive and repulsive polaron resonances
in Fig.~3D yield effective corresponding g-factors of $g_{att-pol} =
18$ and $g_{rep-pol} = -7.2$. This drastic change in the effective
g-factors of elementary optical excitations as compared to the $n_e
= 0$ case depicted in Fig.~3C is a direct consequence of
aforementioned interaction and phase-space filling effects: in the
limit $n_e(-K) \gg n_e(K)$, the  red (blue) shift of the attractive
(repulsive) polaron energy $\Delta_{\pm}^{att} \propto n_e(\mp K)$
($\Delta_{\pm}^{rep} \propto n_e(\mp K)$) stemming from
exciton-electron interactions~\cite{SidlerArxiv2016} is larger for
$\sigma^+$ excitation. On the other hand, the blue-shift due to
phase-space filling (also $\propto n_e(\mp K)$) is more significant
for $\sigma^-$ resonances. For the attractive polaron the two
contributions add, leading to a large splitting. For the repulsive
polaron on the other hand, the interaction and phase-space filling
contributions compete, resulting in an eventual sign change of
$g_{rep-pol}$. These observations provide a further confirmation of
the Fermi-polaron model of excitonic
excitations~\cite{SidlerArxiv2016}.

Figure~3E shows the $\sigma^+$ and $\sigma^-$ PL together with the
differential reflection spectrum at $V_g=-13$~V where both valleys
have $ n_e > 3 \times 10^{12}$~$cm{^{-2}}$. In addition to the
sizeable resonance energy differences, we observe that the
attractive polaron and trion g-factors are not identical
($g_{att-pol} = 14.4$ and $g_{trion} = 13.0$). These differences
provide yet another proof that the elementary excitations
determining absorption/reflection measurements are different from
those that are relevant for PL.


\vspace{5 mm}


Our experiments establish that using moderate $B_z$, it is possible
to valley polarize electron densities exceeding $1.6 \times
10^{12}$~cm$^{-2}$ in a TMD monolayer. This remarkable observation
points to a giant magnetic susceptibility, presumably stemming from
exchange interactions, enabling new possibilities for the control
and manipulation of the valley degree of freedom. Interesting open
questions include the temperature dependence of the magnetic
response and its possible relation to the theoretical prediction of
spontaneous valley polarization in silicon inversion
layers~\cite{ShamPRL1979}. In parallel, our measurements
unequivocally demonstrate the validity of the polaron picture for
describing resonant absorption/reflection experiments. Finally, the
enhancement of the total detected PL intensity by a factor of
$\simeq 8$ for $n_e \sim 1.0 \times 10^{12}$~cm$^{-2}$ suggests a
possible way to increase the radiative quantum efficiency of
MoSe$_2$.



\subsubsection*{Acknowledgments}
The authors also acknowledge many insightful discussions with E.
Demler, A. Kormanyos and G. Burkard. This work is supported by an
ERC Advanced investigator grant (POLTDES) and NCCR QSIT. \\

\clearpage

\begin{figure*}[h!] 
\centering
\includegraphics[scale=0.7]{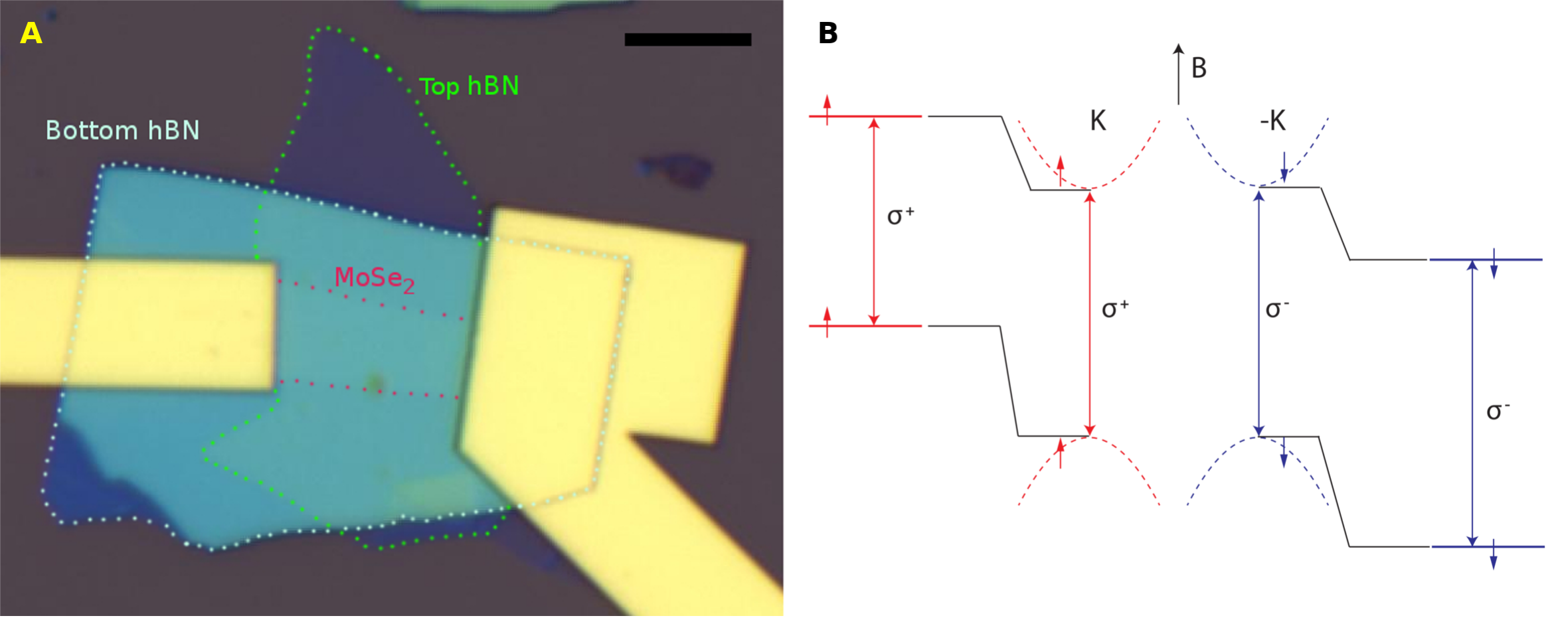}
\caption{ \textbf{A gate controlled MoSe$_2$/hBN heterostructure
under external magnetic field.} \textbf{(A)}, The sample consists of
a $9 \mu$m by $4 \mu$m MoSe$_2$ monolayer sandwiched between two hBN
layers. The heterostructure is placed on top of a $285$~nm thick
SiO$_2$ layer, which in turn is on top of highly doped Si substrate.
A gate voltage applied between the gold contacts to the MoSe$_2$
layer and the highly doped Si allows for controlling the electron
density in the monolayer. \textbf{(B)}, The single particle picture
of conduction and valence band shifts under a magnetic field $B_z$
applied perpendicular to the plane of MoSe$_2$. Assuming a large
spin-orbit splitting, only the lowest (highest) energy conduction
(valence) band is depicted. The contributions from spin,
intra-cellular and inter-cellular currents ensure that for $B_z >
0$, the exciton resonance in the K valley red-shifts along with the
conduction band minimum in the -K valley. This opposite sign of the
effective g-factor of the excitons and electrons plays a key role in
determining the absorption spectrum of MoSe$_2$ monolayers.}
\label{fig1}
\end{figure*}

\clearpage

\begin{figure*}[h!] 
\centering
\includegraphics[scale=0.7]{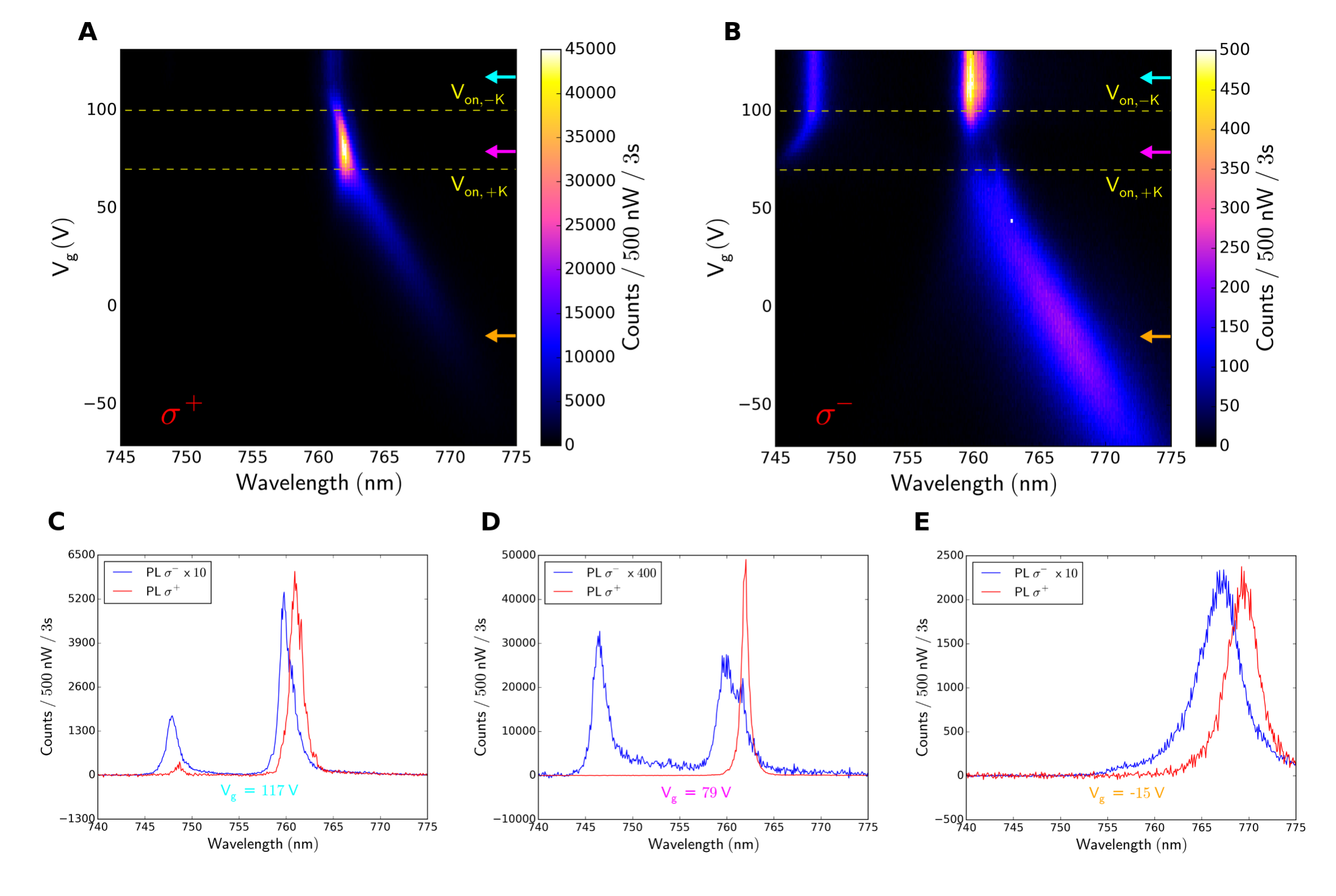}
\caption{ \textbf{Gate voltage dependent photoluminecence spectrum
under moderate magnetic fields.} \textbf{(A)}, Gate voltage ($V_g$)
dependent right-hand  circularly polarized ($\sigma^+$)
photoluminescence (PL) spectrum of a MoSe$_2$/hBN heterostructure at
$B_z = 7$~Tesla under excitation by a linearly polarized $719$~nm
laser. The depicted $V_g$ axis is shifted by $10$~V to compensate
for the hysteretic behavior we observe in the gate scans. \textbf{(B)}, The corresponding PL
spectrum for left-hand circularly polarized ($\sigma^-$).
\textbf{(C)}, The line cut through the $\sigma^+$ and $\sigma^-$ PL
spectra at $V_g = 117$~V~$> V_{on,-K}$ show that the PL exhibits
sizeable degree of polarization where the ratio of $\sigma^+$ and
$\sigma^-$ polarized PL intensities is $\simeq 11$. \textbf{(D)},
The line cut through the $\sigma^+$ and $\sigma^-$ PL spectra at
$V_g = 79$~V shows that the degree of PL polarization increases
dramatically to yield a ratio of $\sigma^+$ and $\sigma^-$ polarized
PL intensities $\simeq 700$. \textbf{(E)}, The line cut through the
$\sigma^+$ and $\sigma^-$ PL spectra at $V_g = -15$~V where both
valleys have high $n_e$. The ratio of the right ($\sigma^+$) and
left ($\sigma^-$) hand circularly polarized PL intensities is
reduced back to $\simeq 11$.} \label{fig2}
\end{figure*}

\clearpage

\begin{figure*}[h!] 
\centering
\includegraphics[scale=0.7]{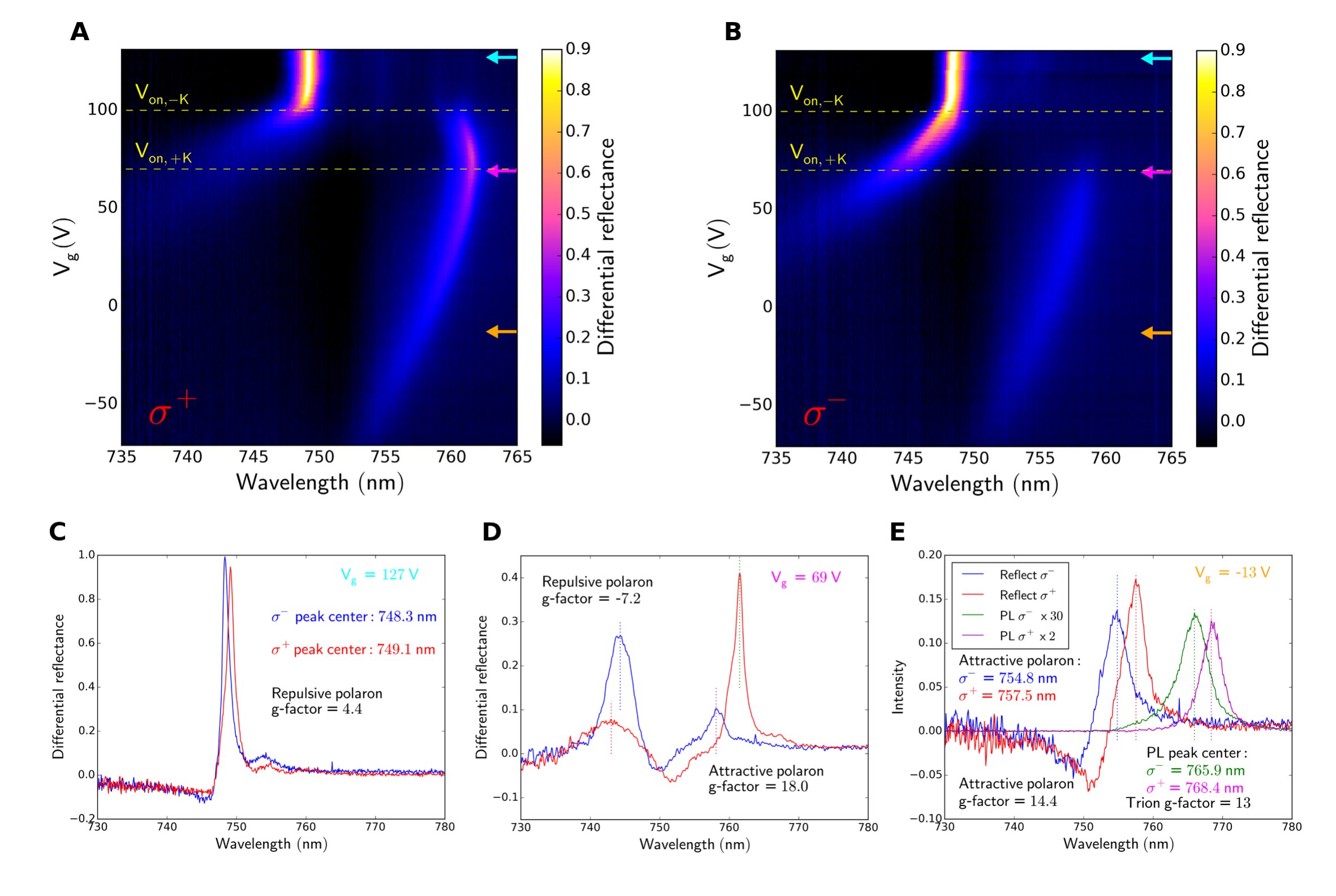}
\caption{ \textbf{Polarization resolved reflection spectrum under
moderate magnetic fields.} \textbf{(A)}, Gate voltage ($V_g$)
dependent right-hand circularly  polarized ($\sigma^+$) white light
reflection spectrum of the MoSe$_2$/hBN heterostructure at a
magnetic field of $B_z = 7$~Tesla.  For $V_g \le V_{on,-K} = 100$~V
(yellow dashed horizontal line), we observe a blue-shift of the
exciton resonance whereas the strength of the attractive polaron
resonance increases as it red-shifts. \textbf{(B)}, Gate voltage
dependent reflection spectrum as in (a) but now for left-hand
circularly polarized ($\sigma^-$) light. For $V_g \le V_{on,-K}$,
the exciton line starts to blue-shift. Only for $V_g \le V_{on,K} =
70$~V, oscillator strength transfer to the attractive polaron is
observed. Whereas the exciton oscillator strength of the $\sigma^+$
and $\sigma^-$ transitions for $V_g > V_{on,-K}$ are nearly
identical, the $\sigma^+$ attractive polaron is much stronger than
its $\sigma^-$ counterpart for $V_{on,K} < V_g < V_{on,-K}$.
\textbf{(C)}, The differential reflection spectrum of both
$\sigma^+$ and $\sigma^-$ light at $V_g = 127$~V: the two resonances
have nearly identical shape and strength but their energies differ
by $1.8$~meV, yielding a g-factor of $4.4$. \textbf{(D)}, The
differential reflection spectrum as in (C) but now at $V_g = 69$~V:
the splitting between the attractive polaron resonances is
$7.3$~meV, which corresponds to a g-factor of $\simeq 18$. The
$\sigma^+$ exciton/respulsive polaron energy on the other hand is
lower than that of $\sigma^-$ by $\simeq 2.9$~meV, yielding a
g-factor of $-7.2$. These results demonstrate that exciton-electron
interactions strongly modify the magneto-optical response of
MoSe$_2$ monolayers. \textbf{(E)}, The differential reflection as
well as PL spectrum  of both $\sigma^+$ and $\sigma^-$ light at $V_g
= -13$~V: both the energy and the g-factor of the resonances
observed in reflection/absorption and PL are different.}
\label{fig3}
\end{figure*}





\begin{thebibliography}{10}

\bibitem{RadisavljevicNNano2011}
B.~Radisavljevic, A.~Radenovic, J.~Brivio, V.~Giacometti, and A.~Kis.
\newblock Single-layer {M}o{S}$_2$ transistors.
\newblock {\em Nat. Nanotechnol.}, 6(3):147--150, march 2011.

\bibitem{SplendianiNanoLett2010}
A.~Splendiani, L.~Sun, Y.~Zhang, T.~Li, J.~Kim, C.-Y. Chim, G.~Galli, and
  F.~Wang.
\newblock Emerging photoluminescence in monolayer {M}o{S}$_2$.
\newblock {\em Nano Lett.}, 10(4):1271--1275, 2010.
\newblock PMID: 20229981.

\bibitem{BaugherNNano2014}
B.~W.~H. Baugher, H.~O.~H. Churchill, Y.~Yang, and P.~Jarillo-Herrero.
\newblock Optoelectronic devices based on electronically tunable p-n diodes in
  a monolayer dichalcogenide.
\newblock {\em Nat. Nanotechnol.}, 9:262--267, Apr 2014.

\bibitem{NovoselovScience2013}
L.~Britnell, R.~M. Bibeiro, A.~Eckmann, R.~Jalil, B.~D. Belle, A.~Mishchenko,
  Y.-J. Kim, R.~V. Gorbachev, T.~Georgiou, S.~V. Morozov, A.~N. Grigorenko,
  A.~K. Geim, C.~Casiraghi, A.~H.~Castro Neto, and K.~S. Novoselov.
\newblock Strong light-matter interactions in heterostructures of atomically
  thin films.
\newblock {\em Science}, 340(6138):1311--1314, 2013.

\bibitem{ChernikovPRL2014}
A.~Chernikov, T.~C. Berkelbach, H.~M. Hill, A.~Rigosi, Y.~Li, O.~B. Aslan,
  D.~R. Reichman, M.~S. Hybertsen, and T.~F. Heinz.
\newblock Exciton binding energy and nonhydrogenic rydberg series in monolayer
  {WS}$_2$.
\newblock {\em Phys. Rev. Lett.}, 113:076802, Aug 2014.

\bibitem{YeNature2014}
Z.~Ye, T.~Cao, K.~O'Brien, H.~Zhu, X.~Yin, Y.~Wang, S.~G. Louie, and X.~Zhang.
\newblock Probing excitonic dark states in single-layer tungsten disulphide.
\newblock {\em Nature}, 513(7517):214--218, September 2014.

\bibitem{HePRL2014}
K.~He, N.~Kumar, L.~Zhao, Z.~Wang, K.~F. Mak, H.~Zhao, and J.~Shan.
\newblock Tightly bound excitons in monolayer {WS}e$_2$.
\newblock {\em Phys. Rev. Lett.}, 113:026803, Jul 2014.

\bibitem{QiuPRL2013}
D.~Y. Qiu, F.~H. da~Jornada, and S.~G. Louie.
\newblock Optical spectrum of {M}o{S}$_2$: Many-body effects and diversity of
  exciton states.
\newblock {\em Phys. Rev. Lett.}, 111:216805, Nov 2013.

\bibitem{WangPRL2014}
G.~Wang, X.~Marie, I.~Gerber, T.~Amand, D.~Lagarde, L.~Bouet, M.~Vidal,
  A.~Balocchi, and B.~Urbaszek.
\newblock Giant enhancement of the optical second-harmonic emission of
  {WS}e$_2$ monolayers by laser excitation at exciton resonances.
\newblock {\em Phys. Rev. Lett.}, 114:097403, Mar 2015.

\bibitem{MakScience2014}
K.~F. Mak, K.~L. McGill, J.~Park, and P.~L. McEuen.
\newblock The valley hall effect in {M}o{S}$_2$ transistors.
\newblock {\em Science}, 344(6191):1489--1492, 2014.

\bibitem{LeeNNano2016}
J.~Lee, K.~F. Mak, and J.~Shan.
\newblock Electrical control of the valley hall effect in bilayer {M}o{S}$_2$
  transistors.
\newblock {\em Nat. Nanotechnol.}, 11(5):421--425, 05 2016.

\bibitem{SrivastavaPRL2015}
A.~Srivastava and A.~Imamoglu.
\newblock Signatures of bloch-band geometry on excitons: Nonhydrogenic spectra
  in transition-metal dichalcogenides.
\newblock {\em Phys. Rev. Lett.}, 115:166802, Oct 2015.

\bibitem{ZhouPRL2015}
J.~Zhou, W.-Y. Shan, W.~Yao, and D.~Xiao.
\newblock Berry phase modification to the energy spectrum of excitons.
\newblock {\em Phys. Rev. Lett.}, 115:166803, Oct 2015.

\bibitem{YeNPhys2016}
Z.~Ye, D.~Sun, and T.~F. Heinz.
\newblock Optical manipulation of valley pseudospin.
\newblock {\em Nat. Phys.}, 2016.

\bibitem{YangNatPhys2015}
L.~Yang, N.~A Sinitsyn, W.~Chen, J.~Yuan, J.~Zhang, J.~Lou, and S.~A Crooker.
\newblock Long-lived nanosecond spin relaxation and spin coherence of electrons
  in monolayer {M}o{S}$_2$ and {WS}$_2$.
\newblock {\em Nat. Phys.}, 11:830--834, 2015.

\bibitem{SanchezArxiv2016}
O.~L. Sanchez, D.~Ovchinnikov, S.~Misra, A.~Allain, and A.~Kis.
\newblock Valley polarization by spin injection in a light-emitting van der
  waals heterojunction.
\newblock {\em Nano Lett.}, 16(9):5792--5797, 2016.

\bibitem{CaoNComm2012}
T.~Cao, G.~Wang, W.~Han, H.~Ye, C.~Zhu, J.~Shi, Q.~Niu, P.~Tan, E.~Wang,
  B.~Liu, and J.~Feng.
\newblock Valley-selective circular dichroism of monolayer molybdenum
  disulphide.
\newblock {\em Nat. Commun.}, 3:887, June 2012.

\bibitem{ZengNNano2012}
H.~Zeng, J.~Dai, W.~Yao, D.~Xiao, and X.~Cui.
\newblock Valley polarization in {M}o{S}$_2$ monolayers by optical pumping.
\newblock {\em Nat. Nanotechnol.}, 7(8):490--493, August 2012.

\bibitem{MakNNano2012}
K.~F. Mak, K.~He, J.~Shan, and T.~F. Heinz.
\newblock Control of valley polarization in monolayer {M}o{S}$_2$ by optical
  helicity.
\newblock {\em Nat. Nanotechnol.}, 7(8):494--498, August 2012.

\bibitem{Durnev-arxiv2016}
D.~Rybkovskiy, Gerber I., and M.~Durnev.
\newblock {https://arxiv.org/abs/1610.02695}.
\newblock Oct 2016.

\bibitem{SidlerArxiv2016}
M.~Sidler, P.~Back, O.~Cotlet, A.~Srivastava, T.~Fink, M.~Kroner, E.~Demler,
  and A.~Imamoglu.
\newblock Fermi polaron-polaritons in charge-tunable atomically thin
  semiconductors.
\newblock {\em Nat. Phys.}, October 2016.

\bibitem{Macdonald2016}
D.~Efimkin and A.~MacDonald.
\newblock {https://arxiv.org/abs/1609.06329}.
\newblock Sep 2016.

\bibitem{Kormanyos2016}
A.~Kormanyos.
\newblock {private communication}.
\newblock Oct 2016.

\bibitem{Fonseca2002}
F.~C. Fonseca, G.~F. Goya, R.~F. Jardim, R.~Muccillo, N.~L.~V. Carre\~no,
  E.~Longo, and E.~R. Leite.
\newblock Superparamagnetism and magnetic properties of ni nanoparticles
  embedded in {SiO}$_2$.
\newblock {\em Phys. Rev. B}, 66:104406, Sep 2002.

\bibitem{ShamPRL1979}
W.~Bloss, L.~Sham, and V.~Vinter.
\newblock Interaction-induced transition at low densities in silicon inversion
  layer.
\newblock {\em Phys. Rev. Lett.}, 43:1529, 1979.

\bibitem{MacNeillPRL2015}
D.~MacNeill, C.~Heikes, K.~F. Mak, Z.~Anderson, A.~Korm\'anyos, V.~Z\'olyomi,
  J.~Park, and D.~C. Ralph.
\newblock Breaking of valley degeneracy by magnetic field in monolayer
  {M}o{S}e$_2$.
\newblock {\em Phys. Rev. Lett.}, 114:037401, Jan 2015.

\bibitem{SrivastavaNPhys2015}
A.~Srivastava, M.~Sidler, A.~V. Allain, D.~S. Lembke, A.~Kis, and A.~Imamoglu.
\newblock Valley zeeman effect in elementary optical excitations of monolayer
  {WS}e$_2$.
\newblock {\em Nat. Phys.}, 11(2):141--147, 02 2015.

\bibitem{AivazianNPhys2015}
G.~Aivazian, Z.~Gong, A.~M. Jones, R.-L. Chu, J.~Yan, D.~G. Mandrus, C.~Zhang,
  D.~Cobden, W.~Yao, and X.~Xu.
\newblock Magnetic control of valley pseudospin in monolayer {WS}e$_2$.
\newblock {\em Nat. Phys.}, 11(2):148--152, February 2015.

\bibitem{LiPRL2014}
Y.~Li, J.~Ludwig, T.~Low, A.~Chernikov, X.~Cui, G.~Arefe, Y.~D. Kim, A.~M.
  van~der Zande, A.~Rigosi, H.~M. Hill, S.~H. Kim, J.~Hone, Z.~Li, D.~Smirnov,
  and T.~F. Heinz.
\newblock Valley splitting and polarization by the zeeman effect in monolayer
  {M}o{S}e$_2$.
\newblock {\em Phys. Rev. Lett.}, 113:266804, Dec 2014.

\end{thebibliography}
\end{document}